\documentclass[twocolumn,aps,superscriptaddress,showpacs]{revtex4}

\usepackage{amssymb}
\usepackage{amsmath}
\usepackage{graphicx}
\usepackage[normalem]{ulem}
\usepackage[dvips]{color}

\begin{document}
\title{isospin splitting of nucleon effective mass and shear viscosity of nuclear matter}

\author{Jun Xu}
\email{xujun@sinap.ac.cn} \affiliation{Shanghai Institute of Applied
Physics, Chinese Academy of Sciences, Shanghai 201800, China}
\affiliation{Kavli Institute for Theoretical Physics China, CAS,
Beijing 100190, China}

\begin{abstract}
Based on an improved isospin- and momentum-dependent interaction, we
have studied the qualitative effect of isospin splitting of nucleon effective mass on the specific shear viscosity of neutron-rich nuclear matter from a relaxation time approach. It is seen that for
$m_n^\star>m_p^\star$, the relaxation time of neutrons is smaller
and the neutron flux between flow layers is weaker, leading to a
smaller specific shear viscosity of neutron-rich matter compared to
the case for $m_n^\star<m_p^\star$. The effect is larger in
nuclear matter at higher densities, lower temperatures, and larger
isospin asymmetries, but it doesn't affect much the behavior of the specific shear viscosity near nuclear liquid-gas phase transition.

\end{abstract}


\maketitle

Understanding the basic strong interaction and the properties of
nuclear matter is the main purpose of nuclear physics. The knowledge
of transport properties of the hot dense matter is
important in understanding the dynamics in heavy-ion collision
experiments as well as the properties of proto-neutron stars. The quark-gluon plasma (QGP) produced
in ultra-relativistic heavy-ion collisions is believed to be a
nearly ideal fluid and has a very small specific shear viscosity
$\eta/s$~\cite{Pes05,Maj07,Son11,Sch11}, i.e., the ratio of the shear viscosity $\eta$ to the
entropy density $s$. It has been further found that the $\eta/s$ decreases with increasing
temperature in the hadronic phase while increases with increasing
temperature in QGP, resulting in a minimum value at the temperature
of hadron-quark phase transition~\cite{Cse06,Lac07}. At even lower
temperatures, the $\eta/s$ of nuclear matter with nucleon
degree of freedom has been investigated from the relaxation time
approach~\cite{Dan84,Shi03,Xu11} and transport model
studies~\cite{Li11,Zho12,Zho13,Fan14}. Similar to the behavior near
hadron-quark phase transition, the $\eta/s$ also shows a minimum in
the vicinity of nuclear liquid-gas phase
transition from various
approaches~\cite{Che07b,Xu13,Pal10b,Zho12,Zho13,Fan14}. Since the correlation between the elliptic flow and the specific shear viscosity seems to be a general feature in both relativistic~\cite{Son11} and intermediate-energy heavy-ion collisions~\cite{Zho14}, in the future it might be promising to measure the $\eta/s$ experimentally, meanwhile providing an alternative way of searching for nuclear liquid-gas phase transition in heavy-ion experiments.

In our previous studies, the specific shear viscosity of neutron-rich matter was investigated based on an isospin- and momentum-dependent interaction~\cite{Xu11,Xu13}. Recently, this interaction has been further improved~\cite{ImMDI} by introducing more parameters so that detailed isovector properties can be studied more flexibly. One of the isovector properties is the neutron-proton effective mass splitting, and it becomes recently a hot topic. The interest was inspired by the recent experimental data of double
neutron/proton ratio from the National Superconducting Cyclotron
Laboratory, which seems to favor a smaller neutron effective mass than
proton based on the calculation using an improved quantum molecular
dynamics model~\cite{Cou14}. However, the well-known Lane potential, representing the nuclear symmetry potential, i.e., the difference between the mean-field potential of neutrons and protons, decreases with increasing nucleon energy, leading to a larger neutron effective mass than proton~\cite{Li04}. To explore the possible uncertainty of neutron-proton effective mass splitting on the $\eta/s$  of hot neutron-rich nuclear matter, we extend our study of the specific shear viscosity with the improved isospin- and momentum-dependent interaction (ImMDI) in this brief report.

The single-nucleon mean-field potential of the ImMDI interaction is written as~\cite{ImMDI}
\begin{eqnarray}
U_\tau(\rho ,\delta ,\vec{p}) &=&A_{u}\frac{\rho _{-\tau }}{\rho _{0}}%
+A_{l}\frac{\rho _{\tau }}{\rho _{0}}  \notag \\
&+&B\left(\frac{\rho }{\rho _{0}}\right)^{\sigma }(1-x\delta ^{2})-4\tau x\frac{B}{%
\sigma +1}\frac{\rho ^{\sigma -1}}{\rho _{0}^{\sigma }}\delta \rho
_{-\tau }
\notag \\
&+&\frac{2C_l}{\rho _{0}}\int d^{3}p^{\prime }\frac{f_{\tau }(%
\vec{p}^{\prime })}{1+(\vec{p}-\vec{p}^{\prime })^{2}/\Lambda ^{2}}
\notag \\
&+&\frac{2C_u}{\rho _{0}}\int d^{3}p^{\prime }\frac{f_{-\tau }(%
\vec{p}^{\prime })}{1+(\vec{p}-\vec{p}^{\prime })^{2}/\Lambda ^{2}},
\label{MDIU}
\end{eqnarray}%
where $\tau=1(-1)$ for neutrons (protons) is the isospin index,
$\rho_n$ and $\rho_p$ are number densities of neutrons and protons,
respectively, $\delta=(\rho_n-\rho_p)/\rho$ is the isospin
asymmetry, with $\rho=\rho_n+\rho_p$ being the total number density,
and $f_{\tau }(\vec{p})$ is the phase-space distribution function.
The $x$ parameter is used to mimic the slope parameter of the
symmetry energy at saturation density $\rho_0$, while additional two
parameters $y$ and $z$ are introduced to adjust the symmetry
potential $U_{sym}$ at infinitely large nucleon momentum and the
value of symmetry energy $E_{sym}$ at saturation density,
respectively, and they enter the functional through the relations
\begin{eqnarray}
A_{l}(x,y)&=&A_{l0} + y + x\frac{2B}{\sigma +1},  \label{AlImMDI}\\
A_{u}(x,y)&=&A_{u0} - y - x\frac{2B}{\sigma +1}, \label{AuImMDI}
\end{eqnarray}
\begin{eqnarray}
C_l(y,z)&=&C_{l0} - 2(y-2z)\frac{p^2_{f0}}{\Lambda^2\ln [(4 p^2_{f0} + \Lambda^2)/\Lambda^2]}, \label{ClImMDI}\\
C_u(y,z)&=&C_{u0} + 2(y-2z)\frac{p^2_{f0}}{\Lambda^2\ln
[(4 p^2_{f0} + \Lambda^2)/\Lambda^2]}, \label{CuImMDI}
\end{eqnarray}
where $p_{f0}$ is the nucleon Fermi momentum in symmetric nuclear
matter at saturation density. The values of parameters $A_{l0}$,
$A_{u0}$, $B$, $C_{l0}$, $C_{u0}$, $\Lambda$, and $\sigma$ as well
as the corresponding macroscopic quantities of nuclear matter from
the ImMDI interaction can be found in Ref.~\cite{ImMDI}.

\begin{figure}[h]
\centerline{\includegraphics[scale=0.8]{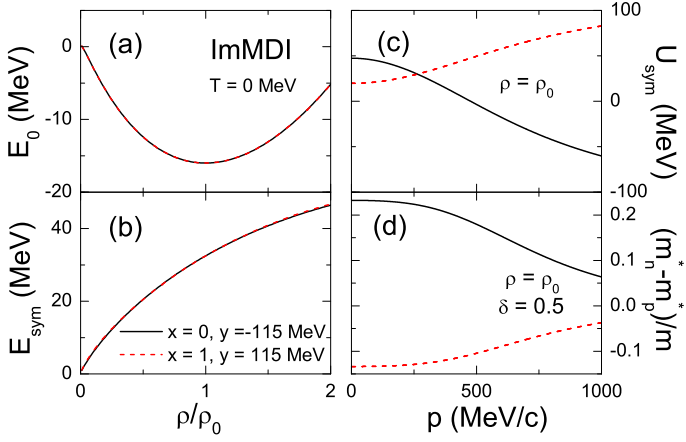}} \caption{(Color
online) Binding energy in symmetric nuclear matter (a), symmetry
energy (b), symmetry potential at saturation density (c), and
relative neutron-proton effective mass splitting (d) in nuclear matter
at saturation density and isospin asymmetry $\delta=0.5$ from
the two parameter sets based on the ImMDI interaction.}
\label{EOSspp}
\end{figure}

The values of $x$, $y$, and $z$ do not affect the isoscalar
properties of nuclear matter, and the binding energy in symmetric
nuclear matter is shown in panel (a) of Fig.~\ref{EOSspp}. The
parameter $z$ is set to be $0$ in the present study, while $x$ and
$y$ change respectively the magnitude and the momentum dependence of
the symmetry potential, and they both contribute to the density
dependence of the symmetry energy. With different
combinations of $x$ and $y$, one can get very similar density
dependence of symmetry energy but different momentum dependence of
the symmetry potential, or equivalently, the isospin splittings of nucleon
effective mass, as can be seen from panels (b), (c), and (d) in
Fig.~\ref{EOSspp}, with the effective mass calculated from
\begin{equation}
\frac{m_{\tau }^{\ast }}{m}=\left( 1+\frac{m}{p}\frac{dU_{\tau
}}{dp}\right) ^{-1}. \label{Meff}
\end{equation}%
The parameter sets [($x=0$), ($y=-115$ MeV)] with
$m_n^\star>m_p^\star$ and [($x=1$), ($y=115$ MeV)] with
$m_n^\star<m_p^\star$ are thus chosen in the following study.

We now briefly review the main ingredient of the relaxation time approach used in previous studies~\cite{Xu11,Xu13}. The shear viscosity is calculated by
assuming that in the uniform nuclear system there exists a static flow field
in the $z$ direction with flow gradient in the $x$ direction. The
shear force, which is related to the nucleon flux as well as the momentum
exchange between flow layers, is proportional to the flow gradient,
and the proportionality coefficient, i.e., the shear viscosity, turns
out to be~\cite{Xu11}
\begin{eqnarray}\label{eta}
\eta = \sum_\tau -\frac{d}{(2\pi)^3} \int \tau_\tau(p) \frac{p_z^2
p_x^2}{pm^\star_\tau} \frac{d n_\tau}{dp} dp_x dp_y dp_z.
\end{eqnarray}
In the above, $d=2$ is the spin degeneracy,
$p=\sqrt{p_x^2+p_y^2+p_z^2}$ is the nucleon momentum, and $n_\tau$ is
the local momentum distribution
\begin{equation}
n_\tau(\vec{p}) = f_\tau(\vec{p})/d = \frac{1}{\exp \left[(\frac{p^{2}}{%
2m}+U_{\tau }(\vec{p})-\mu_\tau )/T\right]+1}
\end{equation}
with $\mu_\tau$ and $T$ being the chemical potential and the
temperature, respectively.  $p_x/m_\tau^\star$ is the nucleon
velocity between flow layers. $\tau_\tau(p)$ is the relaxation time
for a nucleon with isospin $\tau$ and momentum $p$, and it can be
further expressed as
\begin{equation}
\frac{1}{\tau_\tau(p)} = \frac{1}{\tau_\tau^{same}(p)} +
\frac{1}{\tau_\tau^{diff}(p)}, \label{totaltau}
\end{equation}
with $\tau_\tau^{same(diff)}(p)$ being the average collision time
for a nucleon with isospin $\tau$ and momentum $p$ when colliding
with other nucleons of same (different) isospin. For more detailed
derivations and the expressions of the relaxation time, we refer the
readers to Ref.~\cite{Xu11}. The relaxation time depends not only on
the medium properties such as the density, temperature, and isospin
asymmetry, but on the nucleon-nucleon scattering cross section as
well. The free-space proton-proton and neutron-proton cross sections
($\sigma_{NN}^{}$) are taken as the parameterized forms from
Ref.~\cite{Cha90}, while the in-medium cross section is modified by
the effective mass through~\cite{Li05}
\begin{equation}
\sigma^{medium}_{NN} =
\sigma^{}_{NN}\left(\frac{\mu_{NN}^\star}{\mu^{}_{NN}}\right)^2,
\end{equation}
where $\mu_{NN}^{}$ ($\mu_{NN}^\star$) is the free-space (in-medium)
reduced mass of colliding nucleons. The reduced mass scaling of the
in-medium cross section comes from the fact that the differential
cross section is inversely proportional to the relative velocity
between the two colliding nucleons~\cite{Pan92}, while the
difference between the scattering $T$ matrix in free space and in
the nuclear medium is neglected in the present qualitative study.

\begin{figure}[h]
\centerline{\includegraphics[scale=0.8]{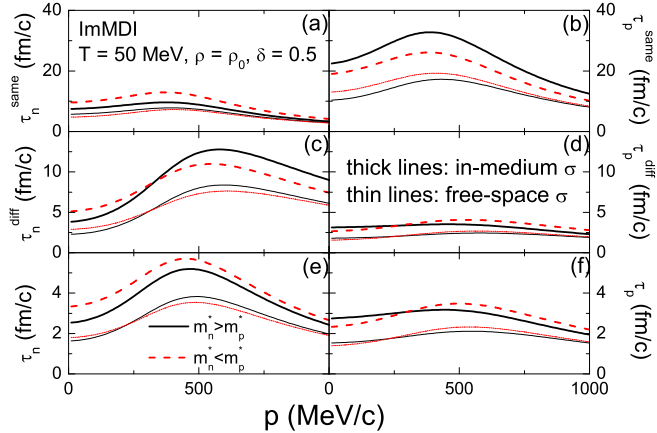}} \caption{(Color
online) Momentum dependence of the total relaxation time and that
for a nucleon to collide with other ones of same or different
isospin in nuclear matter at saturation density and isospin
asymmetry of $\delta=0.5$ using free-space or in-medium nucleon-nucleon scattering cross sections.} \label{tau}
\end{figure}

Figure~\ref{tau} displays the total relaxation time and those for
nucleon scatterings with same or different isospins. As expected,
all relaxation times are larger with in-medium
cross sections compared to the results with those in free space. In neutron-rich nuclear matter, the
scatterings are more frequent for neutron-neutron than for
proton-proton, and protons have more chance to collide with nucleons
of a different isospin than neutrons. The total neutron relaxation
time dominates the shear viscosity due to the sharper neutron
momentum distribution in neutron-rich nuclear matter as can be seen
from Eq.~(\ref{eta}), and it is larger for $m_n^\star<m_p^\star$
than for $m_n^\star>m_p^\star$ as a result of the isospin-dependent
modification from the in-medium effective mass, while the difference
of the total neutron relaxation time for different isospin effective
mass splittings is smaller with free-space cross sections.

\begin{figure}[h]
\centerline{\includegraphics[scale=0.8]{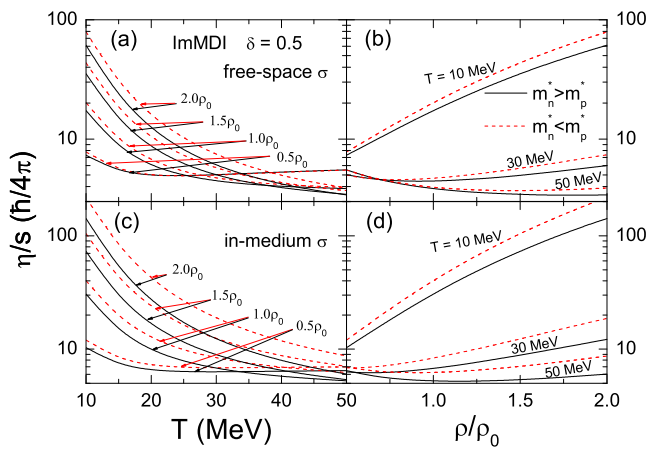}}
\caption{(Color online) Specific shear viscosity with free-space
((a), (b)) and in-medium ((c), (d)) cross sections in nuclear matter of
isospin asymmetry $\delta=0.5$ at different densities and
temperatures for $m_n^\star>m_p^\star$ and $m_n^\star<m_p^\star$.} \label{extensive}
\end{figure}

The extensive results of the specific shear viscosity in nuclear
matter of isospin asymmetry $\delta=0.5$ at various densities and
temperatures are shown in Fig.~\ref{extensive}. Although the
momentum occupation probability $n_\tau(\vec{p})$ at finite
temperature depends on the neutron-proton effective mass splitting~\cite{ImMDI}, the difference of
the entropy density, which is calculated from
\begin{equation}
s =-\sum_\tau d\int [n_{\tau }\ln n_{\tau }+(1-n_{\tau })\ln
(1-n_{\tau })]\frac{d^3p}{(2\pi)^3}, \label{S}
\end{equation}%
turns out to be small between $m_n^\star>m_p^\star$ and
$m_n^\star<m_p^\star$. Despite of the similar relaxation time for
different isospin splittings of nucleon effective mass using
free-space cross sections, a larger neutron effective mass leads to
smaller neutron fluxes between flow layers, reducing the shear
viscosity as can be seen from Eq.~(\ref{eta}). We note that this is a robust feature even if we use a naive mean-free-path formular. The isospin-dependent
modification for the in-medium cross sections further enhances the
difference of the specific shear viscosity between
$m_n^\star>m_p^\star$ and $m_n^\star<m_p^\star$ by giving a smaller
relaxation time for neutrons in the former case, as discussed in
Fig.~\ref{tau}. In addition, one sees that the difference is larger
at higher densities and lower temperatures when the relative isospin
splitting of nucleon effective mass is generally
stronger~\cite{ImMDI}.

\begin{figure}[h]
\centerline{\includegraphics[scale=0.8]{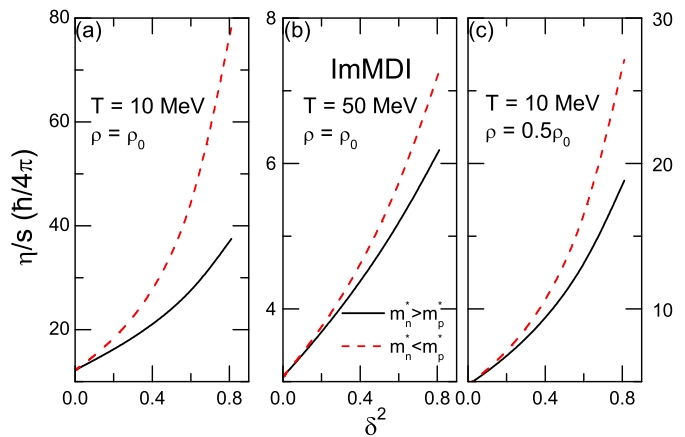}} \caption{(Color
online) Isospin asymmetry dependence of the specific shear viscosity
in nuclear matter at different temperatures and densities for $m_n^\star>m_p^\star$ and $m_n^\star<m_p^\star$. } \label{PA}
\end{figure}

The dependence of the specific shear viscosity on the isospin
asymmetry is displayed in Fig.~\ref{PA} at different temperatures
and densities. It is seen that $\eta/s$ increases with increasing
isospin asymmetry $\delta$ faster than a parabolic relation, and it
increases even faster at higher densities or lower temperatures. One thus expects that the effect discussed here is quite relevant for the evolution of hot neutron stars with a large neutron excess.

\begin{figure}[h]
\centerline{\includegraphics[scale=0.8]{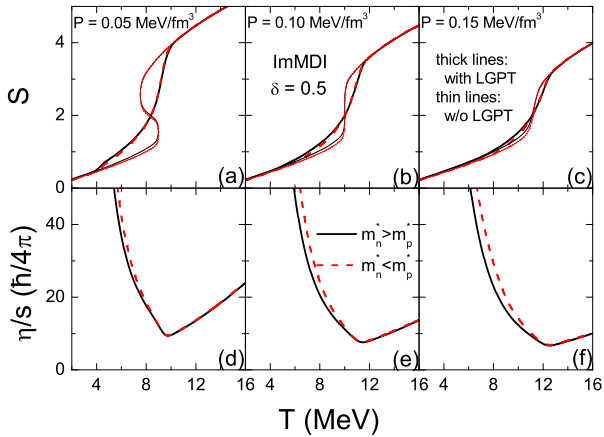}}
\caption{(Color online) Temperature evolution of the entropy (upper
panels) and specific shear viscosity (lower panels) in the presence
of nuclear liquid-gas phase transition (LGPT) at fixed external pressure
$P=0.05$ ((a), (d)), $0.10$ ((b), (e)), and $0.15$ MeV/fm$^3$ ((c),
(f)) and isospin asymmetry $\delta=0.5$ for $m_n^\star>m_p^\star$ and $m_n^\star<m_p^\star$.} \label{etas_LGPT}
\end{figure}

During the liquid-gas phase transition (LGPT) in nuclear matter, each phase satisfying the Gibbs condition~\cite{Mul95,Xu07b} has its own volume fraction. The thin lines in panels (a), (b), and (c) of Fig.~\ref{etas_LGPT}
show the evolution of the entropy when the nuclear matter of isospin
asymmetry $\delta=0.5$ is heated at different fixed pressures. If
the occurrence of the nuclear LGPT is taken
into account, the entropy evolution will follow the thick lines,
with the overall entropy density from that in each phase weighted by
the volume fraction. In infinite nuclear matter the total shear
viscosity can also be calculated from that in each phase weighted by
the volume fraction~\cite{Xu13}, and the temperature evolution of
the specific shear viscosity is shown in panels (d), (e), and (f) of
Fig.~\ref{etas_LGPT}. A minimum value of $\eta/s$ is seen at a
higher temperature with increasing pressure, and this value is also
smaller at a larger external pressure. The isospin splitting of
nucleon effective mass has very small effects on the entropy
evolution, while the specific shear viscosity is smaller for
$m_n^\star>m_p^\star$ than for $m_n^\star<m_p^\star$ at lower
temperatures (higher densities) in the liquid phase side, but the
difference is negligible at higher temperatures (lower densities) in
the gas phase side. Moreover, the minimum point of $\eta/s$ is not affected by
the isospin splitting of the nucleon effective mass. This general feature, which is not sensitive to detailed nuclear interaction, might be helpful in searching for the occurrence of the nuclear LGPT in low- and intermediate-energy heavy-ion
collisions if people find ways to measure the specific shear
viscosity there~\cite{Zho14}.

In summary, the specific shear viscosity with different isospin
splittings of nucleon effective mass has been studied in
neutron-rich nuclear matter based on an improved isospin- and
momentum-dependent interaction. Qualitatively, it is seen that the specific shear
viscosity is larger for $m_n^\star<m_p^\star$ than for
$m_n^\star>m_p^\star$, and the difference is more obvious at higher
densities, lower temperatures, and larger isospin asymmetries. This
is due to different neutron fluxes between flow layers as well as
the isospin-dependent modification to the in-medium nucleon-nucleon
cross sections. On the other hand, the behavior of the specific
shear viscosity near nuclear liquid-gas phase transition remains robust and seems to be insensitive to the detailed nuclear interaction. Our study may be helpful in
understanding the transport properties of the hot neutron-rich
nuclear matter produced in heavy-ion collision experiments as well
as that in hot neutron stars.

This work was supported by the Major State Basic Research
Development Program (973 Program) in China under Contract Nos.
2015CB856904 and 2014CB845401, the National Natural Science
Foundation of China under Grant Nos. 11475243 and 11421505, the "100-talent plan"
of Shanghai Institute of Applied Physics under Grant No. Y290061011
from the Chinese Academy of Sciences, and the "Shanghai Pujiang
Program" under Grant No. 13PJ1410600.

\end{document}